\documentclass[a4paper, 12pt, authoryear]{elsarticle}

\usepackage{amsmath}
\usepackage{amssymb}
\usepackage{amsthm}

\usepackage{lineno}

\usepackage{graphicx}
\usepackage{epstopdf}
\usepackage{multirow}
\usepackage{booktabs}
\usepackage{lipsum}
\usepackage{hyperref}
\hypersetup
{%bookmarks=true,     % show bookmarks bar?
 unicode=false,     % non-Latin characters in Acrobat's bookmarks
 pdftoolbar=true,    % show Acrobat's toolbar?
 pdfmenubar=true,    % show Acrobat's menu?
 pdffitwindow=false,   % window fit to page when opened
 pdfstartview={FitH},  % fits the width of the page to the window
 pdftitle={My title},  % title
 pdfauthor={Author},   % author
 pdfsubject={Subject},  % subject of the document
 pdfcreator={Creator},  % creator of the document
 pdfproducer={Producer}, % producer of the document
 pdfkeywords={keywords}, % list of keywords
 pdfnewwindow=true,   % links in new window
 colorlinks=true,    % false: boxed links; true: colored links
 linkcolor=red,     % color of internal links
 citecolor=blue,    % color of links to bibliography
 filecolor=magenta,   % color of file links
 urlcolor=cyan      % color of external links
}

\usepackage{subcaption}
\usepackage{placeins}

\usepackage{siunitx}
\usepackage{nomencl}

\makenomenclature

\usepackage{etoolbox}
\renewcommand\nomgroup[1]{%
  \item[\bfseries
  \ifstrequal{#1}{A}{Sets}{%
  \ifstrequal{#1}{B}{Functions}{%
  \ifstrequal{#1}{C}{Parameters}{%
  \ifstrequal{#1}{D}{Variables}{}}}}%
]}

\newcommand{\nomunit}[1]{%
\renewcommand{\nomentryend}{\hspace*{\fill}#1}}

\journal{-}

\begin{document}

\begin{frontmatter}



\title{Electric Vehicle E-hailing Fleet Dispatching and Charge Scheduling}


\author[USYD]{Linji Chen}
\author[RMIT]{Homayoun Hamedmoghadam}
\author[RMIT]{Mahdi Jalili}
\author[USYD]{Mohsen Ramezani*}\ead{mohsen.ramezani@sydney.edu.au}
\address[USYD]{The University of Sydney, School of Civil Engineering, Australia}
\address[RMIT]{RMIT University, School of Engineering, Australia}

% TODO: Add results in abstract
\begin{abstract}
    With recent developments in vehicle and battery technologies, electric vehicles (EVs) are rapidly getting established as a sustainable alternative to traditional fossil-fuel vehicles. This has made the large-scale electrification of ride-sourcing operations a practical viability, providing an opportunity for a leap toward urban sustainability goals. Despite having a similar driving range to fossil-fuel vehicles, EVs are disadvantaged by their long charging times which compromises the total fleet service time. To efficiently manage an EV fleet, the operator needs to address the charge scheduling problem as part of the dispatch strategy. This paper introduces a probabilistic matching method which evaluates the optimal trip and charging decisions for a fully electrified e-hailing fleet, with the goal of maximising the operator's expected market profit. In the midst of the technological transition towards autonomous vehicles, it is also critical to include stochastic driver behaviours in transport models as presented in this paper. Since drivers may either comply with trip dispatching or choose to reject a charging trip order considering the additional fees, contrary to the commonly assumed fleet autonomy, the proposed method designs an incentivisation scheme (charging discounts) to encourage driver compliance so that the planned charging trips and the associated profit can be realised.
\end{abstract}


% \begin{highlights}
% \end{highlights}


\end{frontmatter}

%\linenumbers

\newpage
\section{Introduction}
\label{sec:introduction}
    In recent years, on-demand point-to-point trip services have become an essential pillar in urban transportation. Equipped with new technologies, transportation network companies (TNCs) have the potential to make an impact on both the market and the environment. One of the new opportunities arises from the increasing adoption rate of electric vehicles (EVs). With continuing advances in vehicle and battery technologies, as well as more investments in public charging infrastructures, EVs have become a viable mode for ride-sourcing services. As a result, companies would need efficient management plans for their EV fleet to address problems such as charge scheduling and energy utilisation. An efficient fleet dispatch method can be beneficial in terms of both the profit and the battery usage. By balancing the fleet state of charge (SoC) with the dynamic spatio-temporal demand levels, the company would be able to serve more customers.
    
    The EV fleet management problem has been investigated in recent studies from the engineering perspective. \citet{yi_framework_2021} proposes an optimisation method to centrally plan dispatch and charging actions for a fleet of autonomous electric vehicles (AEVs). The optimisation is able to achieve more trip deliveries than the heuristic benchmark strategy. The management problem is often formulated as a Markov decision process which is subsequently optimised by either reinforcement learning \cite{shi_operating_2020}, neural networks \cite{kullman_dynamic_2021, yu_optimal_2021}, or dynamic programming \cite{al-kanj_approximate_2020} methods. These methods are dependent on reliable value function approximations which often require extensive data collection. Some literature generalise the problem as a dial-a-ride problem for EVs. \citet{bongiovanni_electric_2019} formulates it as a mixed-integer linear problem and presents a branch-and-cut algorithm to solve small-scale instances. \citet{iacobucci_optimization_2019} applies model predictive control methods at two time aggregation levels to derive the optimal decisions.
    
    An extensive catalogue of literature investigate the fleet management problem for traditional vehicles under different assumptions. Most solutions dispatch the moving agents to stationary resources in a centralised manner. \citet{wong_cell-based_2014, wong_two-stage_2015} consider a cell-based network and recommend the most profitable search path for taxis based on the cumulative probability of finding passenger pick-ups. With regard to a dynamic demand and supply environment, \citet{ramezani_dynamic_2018} applies macroscopic approaches to evaluate the optimal repositioning decisions for the taxi fleet in a large city network. In \citet{duan_centralized_2020}, the centralised dispatching system is combined with decentralised autonomous taxis to distribute the computational workload. In recent years, new management methods have been developed for autonomous vehicle (AV) fleets \cite{horl_fleet_2019, vosooghi_shared_2019, hyland_dynamic_2018, ma_designing_2017}. The AV technology can reinforce the feasibility of new energy vehicles by reducing the operational uncertainties. For interested readers, \citet{narayanan_shared_2020} provides review on shared AV services.
    
    However, most literature assume full autonomy or driver compliance, meaning that EVs would follow the instructions even with little benefits. To ensure successful execution of the optimal strategy, an incentivisation policy should be designed. This paper uses a probabilistic matching method to centrally dispatch a fleet of EVs to waiting passengers and available chargers in the network. The proposed matching method aims to maximise the expected market profit, considering the future profitability. Another contribution of this paper is the design of an incentivisation policy to encourage EV drivers to abide by the optimal dispatch orders.
    
    The paper is structured as follows. Section~\ref{sec:problem} defines the problem and general assumptions. Section~\ref{sec:methodology} explains the dispatching problem and probabilistic matching in detail. Section~\ref{sec:behaviour} describes relevant passenger and driver behavioural models that lead to the stochastic and dynamic market conditions. Section~\ref{sec:simulation} elaborates the simulation environment and agent behaviours in detail. Preliminary results are discussed in Section~\ref{sec:result}. Last, Section~\ref{sec:conclusion} summarises current findings and outlines future directions.

\section{Problem Overview}
\label{sec:problem}
    This paper develops a centralised matching method for human-driven EVs in an on-demand mobility market considering the dynamic demand-supply relationships, time-varying charging prices, and stochastic dispatch compliance behaviours of human drivers. The objective is to maximise the TNC's expected profits via optimal batch matching solutions and charging incentives. The matching outcomes for vacant EVs are (1) a waiting passenger, (2) a public EV charger, or (3) no action. Drivers are assumed to always accept a passenger trip for their personal income. However, they do not show absolute compliance when dispatched to recharge their vehicles. Since time and money are consumed when recharging EVs, such trips are less appealing to drivers without financial incentives. The TNC can offer discounted charging prices to improve driver compliance for the optimal matching solution, maximising the expected benefit over a longer period.
    
    Waiting passengers request for trip services between their respective origins and destinations (ODs). Passengers cancel their requests if the waiting time before a successful matching exceeds the passenger's patience (Type I cancellation), or if the matched EV takes too long to pick up the passenger (Type II cancellation). Human drivers may choose their shift hours flexibly to gain personal incomes for passenger trip services. As the vehicle state of charge (SoC) drops, drivers can recharge their EVs at charging stations throughout the network. Charging stations are also operated by the TNC. Thus, a balance between charging profit and trip profit should be achieved.
    
    This paper make the following assumptions:
    
    \begin{enumerate}
        \item The on-demand mobility market is monopolised by a single TNC. The company operates a ride-sourcing fleet which consists of only EVs. 
        \item This paper does not investigate the effects of pricing structures. Both passenger trip fare and driver wage are fixed rates based on the in-vehicle (occupied) trip time.
        \item The TNC also owns or leases public charging stations as a source of profit. The charger locations are pre-determined and fixed.
        \item The road travel times are fixed throughout the day regardless of the congestion levels. EVs follow pre-computed shortest travel time paths for all passenger and charging trips.
        \item Partial charging is not allowed. EVs always recharge to $90$\% of their maximum capacity, which protects batteries from degradation after frequent recharging.
        \item Since each charging station can only serve a limited number of vehicles, it is possible for EVs to queue for recharging. The TNC can accurately predict recharge completion times.
        \item Passengers exhibit realistic trip cancellation behaviours. They can reject trip matchings with low quality of service, e.g., when experiencing long matching time or pick-up time.
    \end{enumerate}
    
    It is crucial to recognise the dynamic nature of the problem. Without any knowledge of passenger demands or the charger availability in advance, the mission planning is challenged by such uncertainties. While the batch matching solution can guarantee optimality at the time of matching, future evolution of the market condition is neglected. In \citet{hyland_dynamic_2018}, assigned vehicles can be re-assigned to pick up different resources and the strategy significantly reduces the total distance travelled. However, re-assignment can cause confusion among human drivers. It also complicates the fare, wage, and incentive calculations. This paper aims to achieve efficient fleet dispatching without revoking past matching decisions.

\section{Methodology of Probabilistic Matching}
\label{sec:methodology}
\begin{figure}[htb!]
        \centering
        \includegraphics[width=0.45\textwidth]{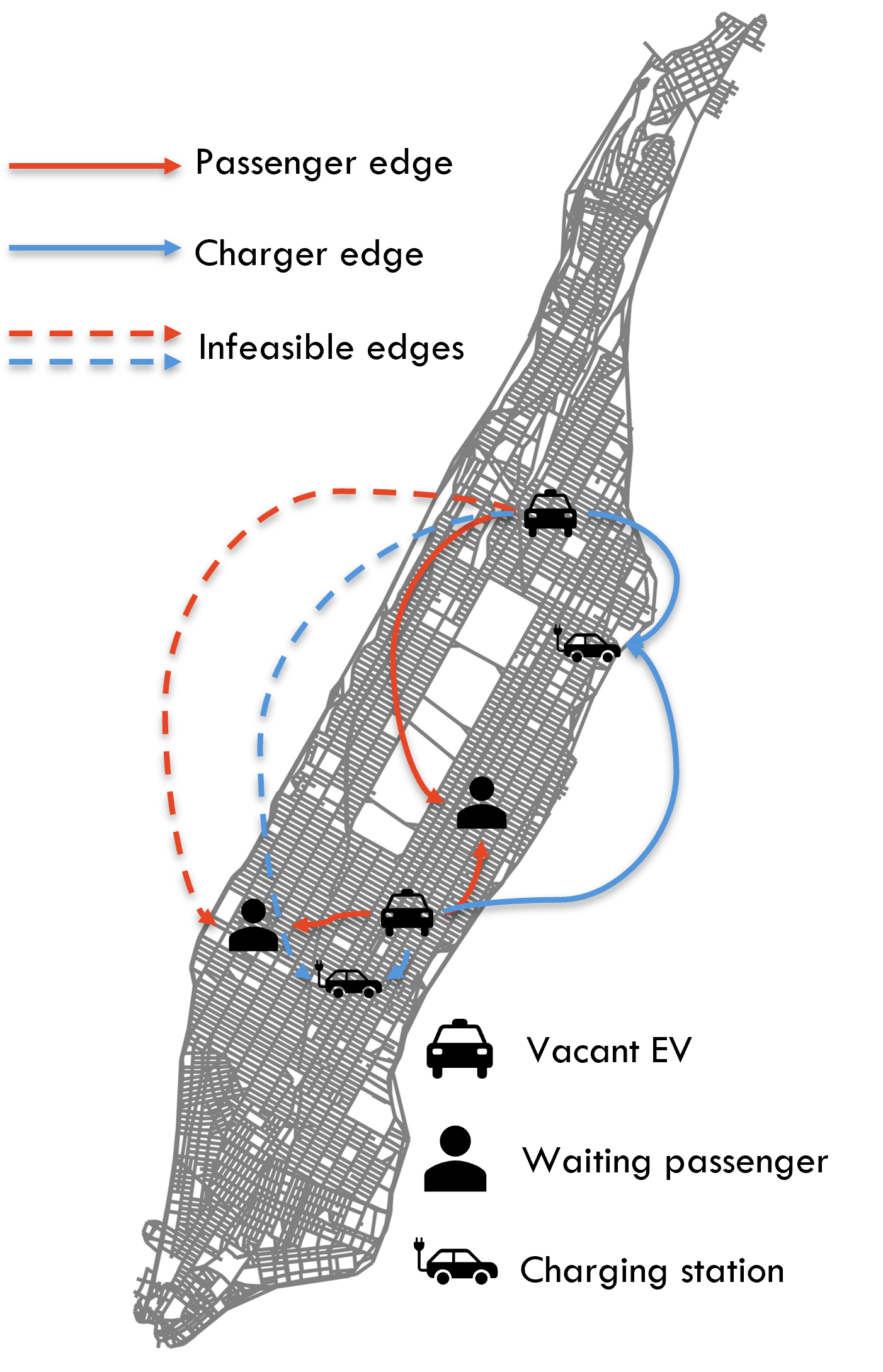}
        \caption{Illustration of the batch matching options. Vacant EVs are matched with waiting passengers and chargers within their SoC travel range.}
        \label{fig:matching_illustration}
    \end{figure}

    % Matching outcomes, batch matching explanation
    An example of feasible and infeasible outcomes for the TNC at a matching instance is illustrated in Figure~\ref{fig:matching_illustration}. At each discrete matching interval (e.g., $10$ seconds), vacant EVs in the network are assigned to feasible waiting passengers or chargers within their travel range (limited by the vehicle SoC). The matching outcome is a set of one-to-one pairs between a vacant vehicle $i$ in the vehicle set $\mathcal{V}$ and a waiting passenger $j$ in the passenger set $\mathcal{P}$, or a charger (available or occupied) $j$ in the charger set $\mathcal{C}$. The optimal matchings are dependent on the expected benefits (see Section~\ref{sec:methodology_weights}), which consider the monetary profits, quality of service measures, and the marginal value of charge (see Section~\ref{sec:value_of_charge}). To represent a realistic scenario, human drivers and passengers react to the matching results based on their behavioural attributes such as passenger's acceptance of pick-up times, and driver's compliance for the dispatched charging trips (See Section~\ref{sec:behaviour}).
    
\nomenclature[A]{$\mathcal{V}(k)$}{Vacant EVs \nomunit{-}}
\nomenclature[A]{$\mathcal{P}(k)$}{Waiting passengers \nomunit{-}}
\nomenclature[A]{$\mathcal{C}$}{Chargers \nomunit{-}}
    
\subsection{Expected benefit of vehicle dispatching actions}
\label{sec:methodology_weights}
    % Weight explanation
    In this paper, the proposed matching method aims to maximise the TNC's total profit. The effectiveness of the proposed matching algorithm is influenced by the TNC's expected benefit for each dispatching option. The optimisation problem identifies $3$ aspects of benefits related to the profit, (1) the monetary passenger trip profit, (2) reduction in unprofitable deadheading time, and (3) potential improvement in the marginal value of charge for each EV, which quantifies the benefit of additional vehicle SoC for future trips in a local neighbourhood.

    % Monetary profits
    For passenger trips, the monetary profit calculation is straightforward. The TNC collects trip fares and pays driver wages based on the in-vehicle occupied time (or passenger on-board time), $t_j^\mathrm{o}$, which is determined by the passenger OD. This paper assumes fixed fare ($f$) and wage ($w$) rates per unit time. Thus, the monetary profit for serving passenger $j$ is $(f - w) t_j^\mathrm{o}$.

\nomenclature[C, 01]{$f$}{Passenger fare rate \nomunit{$\si{\$\per\second}$}}
\nomenclature[C, 01]{$w$}{Driver wage rate \nomunit{$\si{\$\per\second}$}}
\nomenclature[D, 06]{$t_j^\mathrm{o}$}{In-vehicle travel time of passenger $j$, equivalently EV occupied time \nomunit{$\si{\second}$}}

    % Passenger waiting times
    The TNC needs to maintain its reputation in the long term by providing good quality of service. For each matching, this can be achieved by reducing the passenger waiting time. The waiting time contains two parts; time between the trip request and the matching instance ($t_j^\mathrm{m}$), and the pick-up time for the vehicle to reach the matched passenger ($t_{ij}^\mathrm{p}$). Passenger waiting times directly influence their cancellation behaviours (Section~\ref{sec:behaviour_passenger}). A longer $t_{ij}^\mathrm{p}$ reduces the expected benefit due to worse quality of service, which is likely to induce more passenger trip cancellations. On the other hand, passengers who have waited longer for matchings should be given a higher priority to avoid cancellations. Thus, a longer $t_j^\mathrm{m}$ is defined to increase the expected trip benefit to represent the matching urgency.

\nomenclature[D, 05]{$t_j^\mathrm{m}$}{Waiting time of of passenger $j$ since request till a matching instance \nomunit{$\si{\second}$}}
\nomenclature[D, 06]{$t_{ij}^\mathrm{p}$}{Waiting time of passenger $j$ since a successful matching till pick-up, equivalently the travel time from matched EV to passenger trip origin \nomunit{$\si{\second}$}}

    % Consumption of SoC
    Vehicle SoC is consumed to complete passenger trips. The matched EV travels from its current location to pick up the passenger after $t_{ij}^\mathrm{p}$ and delivers the passenger to the destination after the in-vehicle occupied time $t_j^\mathrm{o}$. With a SoC consumption rate of $Q_i^\mathrm{o}$ [kW], each passenger trip would consume $Q_i^\mathrm{o}\left(t_{ij}^\mathrm{p} + t_j^\mathrm{o} \right)$ [kWh].

\nomenclature[C, 04]{$Q_i^\mathrm{o}$}{Constant EV SoC consumption rate of vehicle $i$ \nomunit{$\si{\kilo\watt}$}}
    
    % Charging profits
    The TNC also operates the charging facilities so that the charging prices and discount incentives can be dynamically adjusted. At a time instance $k$, the dynamic charging price for each driver consists of $3$ components: (1) profit margin, $e_1$, which secures the operational benefits for charging services; (2) infrastructure cost, $e_2$, which represents the lease or maintenance fee of the charging facilities; and (3) time-of-use tariff, $e_3(k)$, which differentiates the peak and off-peak hours for the TNC to cover the electricity price. When no incentive is offered, the TNC's charging profit is equal to the profit margin, $e_1$. When the TNC offers full incentive, the profit is in deficient equal to the infrastructure and time-of-use costs, $-\left(e_2 + e_3(k)\right)$. The discount ($r_{ij}$) is bounded between $0\%$ and $100\%$, and is offered by the TNC for each matching option to the chargers. Thus, the discount is dynamic based on the market conditions, and heterogeneous for different matching pairs. The TNC's unit charging profit, $e_{ij}(k)$ [\$/kWh], can be expressed as
    \begin{equation*}
        e_{ij}(k) = e_1 - r_{ij}(k) \cdot \left(e_1 + e_2 + e_3(k)\right).
    \end{equation*}

\nomenclature[C, 00]{$k$}{Time instance \nomunit{-}}
\nomenclature[C, 02]{$e_1, e_2$}{Charging price component, profit margin and infrastructure cost \nomunit{$\si{\$\per\kilo\watt\hour}$}}
\nomenclature[C, 02]{$e_3(k)$}{Charging price component, dynamic time-of-use tariff \nomunit{$\si{\$\per\kilo\watt\hour}$}}
\nomenclature[D, 03]{$r_{ij}$}{Discount incentive offered by the TNC for driver $i$ to charger $j$ \nomunit{-}}
\nomenclature[B, 01]{$e_{ij}$}{Overall charging price paid by driver $i$ at charger $j$ \nomunit{$\si{\$\per\kilo\watt\hour}$}}

    % Charged SoC
    Before charging their EV at the respective charging station, the driver needs to travel to the charger location, consuming some time ($t_{ij}^\mathrm{c}$) and SoC in the process. Upon reaching the charging stations, the EV SoC is raised from its current level at time $k$, $s_i(k)$, to $90$\% of the maximum capacity, $s_i^\mathrm{max}$. With a SoC consumption rate of $Q_i^\mathrm{o}$ per unit of travel time, the recharged SoC amount can be expressed as
    \begin{equation}\label{eqn:charging_amount}
        \Delta s_i = 0.9 s_i^\mathrm{max} - \left(s_i(k) - Q_i^\mathrm{o} t_{ij}^\mathrm{c}\right).
    \end{equation}

\nomenclature[C, 03]{$s_i^\mathrm{max}$}{Maximum capacity of SoC of vehicle $i$ \nomunit{$\si{\kilo\watt\hour}$}}
\nomenclature[D, 01]{$s_i(k)$}{SoC of vehicle $i$ at time instance $k$ \nomunit{$\si{\kilo\watt\hour}$}}
\nomenclature[B, 02]{$\Delta s_i$}{Amount of SoC to be charged for vehicle $i$ \nomunit{$\si{\kilo\watt\hour}$}}

    % Charging time
    For a charging trip, the EV spends time to (i) arrive at the matched charging station ($t_{ij}^\mathrm{c}$), (ii) queue at the charging station for it to become available if occupied by other EVs ($t_{ij}^\mathrm{q}$), and (iii) recharge vehicle SoC to 90\% of the maximum capacity. Note that the queuing time in Equation~\eqref{eqn:charging_queue_time} is measured after the EV arrives at the charging station, which is available after some time instance $k_j^\mathrm{q}$. With a linear charging rate of $Q_{ij}^\mathrm{x}$ [kW], Equation~\eqref{eqn:charging_time} expresses the total time spent to complete a charging trip.
    \begin{align}
        t_{ij}^\mathrm{q} &= \max\left(0, k_j^\mathrm{q} - (k + t_{ij}^\mathrm{c})\right) \label{eqn:charging_queue_time} \\
        \Delta k_{ij}^\mathrm{c} &= t_{ij}^\mathrm{c} + t_{ij}^\mathrm{q} + \Delta s_i / Q_{ij}^\mathrm{x}. \label{eqn:charging_time}
    \end{align}

\nomenclature[D, 05]{$t_{ij}^\mathrm{c}$}{Travel time between EV $i$ and charger $j$ \nomunit{$\si{\second}$}}
\nomenclature[D, 04]{$k_j^\mathrm{q}$}{Available (queue dissipation) time instance of charger $j$ \nomunit{-}}
\nomenclature[B, 03]{$t_{ij}^\mathrm{q}$}{Queuing time of EV $i$ at charger $j$ \nomunit{$\si{\second}$}}
\nomenclature[B, 04]{$\Delta k_{ij}^\mathrm{c}$}{Total charging time of EV $i$ at charger $j$ since matching \nomunit{$\si{\second}$}}
\nomenclature[C, 04]{$Q_{ij}^\mathrm{x}$}{Constant charging speed of EV $i$ at charger $j$ \nomunit{$\si{\kilo\watt}$}}

    % Marginal value of charge
    The marginal value of charge ($\rho_\mathrm{e}$) represents the estimated profit for each additional unit of vehicle SoC in a local area around the EV. As a vehicle completes a passenger trip or a charging trip, its SoC changes over time and impacts the supply side of the future market around the trip destination. The difference between $\rho_\mathrm{e}$ in the current zone $z_i$ at time $k$ and a predicted market in zone $z_j$ at some future time $k'$ (denoted as $\Delta \rho_\mathrm{e}^\mathrm{p}$ or $\Delta \rho_\mathrm{e}^\mathrm{c}$ depending on the trip type) quantifies the predicted improvement in market profitability. The proposed method includes the predicted effects of fleet SoC variations as part of the expected market benefit. Details of $\rho_\mathrm{e}$ are elaborated in Section~\ref{sec:value_of_charge}.
    
    % Value of time
    For simplicity, the values of time are assumed to be constant during the simulated $24$-hour operation. Since different activities are likely to have different impacts on the TNC's expected benefit, three values of time parameters are introduced, (1) $\rho_\mathrm{t}^\mathrm{pm}$ is associated with the passenger waiting time before a successful matching, (2) $\rho_\mathrm{t}^\mathrm{pp}$ is associated with the pick-up waiting time, and (3) $\rho_\mathrm{t}^\mathrm{c}$ is associated with the charging time. In general, $\rho_\mathrm{t}^\mathrm{c}$ takes a lower value than $\rho_\mathrm{t}^\mathrm{pm}$ and $\rho_\mathrm{t}^\mathrm{pp}$ because passenger cancellations lead to reputation loss on top of the revenue loss. The value of $\rho_\mathrm{t}^\mathrm{pm}$ is also higher than $\rho_\mathrm{t}^\mathrm{pp}$ because passengers usually demonstrate higher patience for in-vehicle travel time than the out-of-vehicle waiting time \citep{yan_integrating_2019}.

\nomenclature[C, 05]{$\rho_\mathrm{t}^\mathrm{pm}$}{Value of time for waiting passenger to be matched \nomunit{$\si{\$\per\second}$}}
\nomenclature[C, 05]{$\rho_\mathrm{t}^\mathrm{pp}$}{Value of time for matched passenger to be picked up \nomunit{$\si{\$\per\second}$}}
\nomenclature[C, 05]{$\rho_\mathrm{t}^\mathrm{c}$}{Value of time for EV to complete charging \nomunit{$\si{\$\per\second}$}}

    % Expected benefit summary
    In summary, the TNC evaluates the expected benefit based on the monetary profit, time cost, and the value of SoC consumption or recharging. Equations~\eqref{eqn:benefit_passenger} and \eqref{eqn:benefit_charging} specify the expected benefit in monetary terms for a passenger trip and charging trip, respectively.
    \begin{align}
        \pi_{ij}(k) &= (f - w) t_j^\mathrm{o} + \rho_\mathrm{t}^\mathrm{pm} t_j^\mathrm{m} - \rho_\mathrm{t}^\mathrm{pp} t_{ij}^\mathrm{p} + \Delta \rho_\mathrm{e}^\mathrm{p} Q_i^\mathrm{o}\left(t_{ij}^\mathrm{p} + t_j^\mathrm{o} \right) & \forall j \in \mathcal{P} \label{eqn:benefit_passenger} \\
        \pi_{ij}(k) &= e_{ij}(k) \Delta s_i - \rho_\mathrm{t}^\mathrm{c} \Delta k_{ij}^\mathrm{c} + \Delta \rho_\mathrm{e}^\mathrm{c} \Delta s_i & \forall j \in \mathcal{C} \label{eqn:benefit_charging}
    \end{align}

\nomenclature[D, 02]{$x_{ij}$}{Binary variable representing the matching outcome between a pair \nomunit{-}}
\nomenclature[B, 06]{$\hat{\pi}(z_i, k)$}{Estimated market benefit in the current local market \nomunit{\$}}
\nomenclature[B, 06]{$\hat{\pi}(z_j, k')$}{Estimated market benefit in a future local market \nomunit{\$}}
\nomenclature[B, 06]{$\pi_{ij}(k)$}{Expected matching benefit for the TNC \nomunit{\$}}

\subsection{Expected charging benefit and marginal value of charge}
\label{sec:value_of_charge}
\begin{figure}[htb!]
        \centering
        \includegraphics[width=0.85\linewidth]{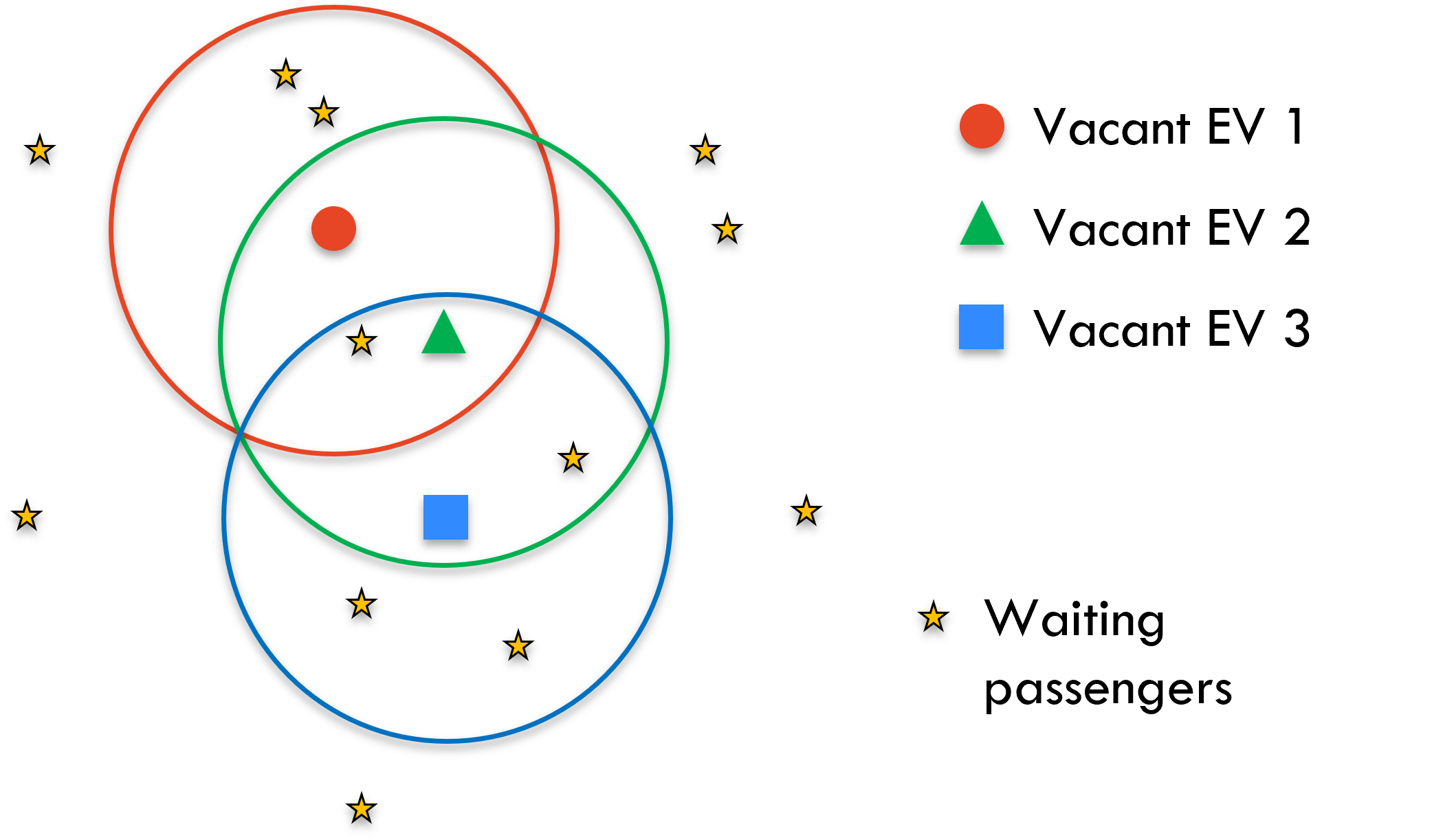}
        \caption{Illustration of zones around vacant EVs in a network. Assuming the travel time is proportional to the Euclidean distance between two points in the network, the circular zones represent the local markets. For example, there are $2$ vacant EVs and $3$ waiting passengers in zone $1$. The zones can be overlapping such that both EV $1$ and EV $2$ are in zone $1$. Note that the mutual inclusion is not always true in a directed graph when the travel times are different in two directions.}
        \label{fig:zones}
    \end{figure}

    Given the regional trip demand-supply heterogeneity in an e-hailing market, this paper defines the marginal value of charge ($\rho_\mathrm{e}$) as a dynamic metric of the expected benefit for the TNC per additional unit of fleet SoC in a local area around each vacant EV. The boundaries of such zones are defined by the travel time from each vacant EV, and may be overlapping so that a vacant EV may appear in multiple zones at the same time. Figure~\ref{fig:zones} explains the concept of the zones representing the local markets around vacant EVs.

    The marginal value of charge ($\rho_\mathrm{e}$) in zone $z_i$ around vehicle $i$ at time $k$ is determined via an algorithm consisting of two steps. First, a local profit baseline is estimated for each zone by performing a batch matching with vacant EVs and waiting passengers in each zone, aiming to minimise the total pick-up waiting time. The matching outcome indicates potential trip profits with the current level of fleet SoC. Then, the matching is repeated without vehicle $i$ to estimate the loss in trip profit at a lower fleet SoC level. As a result, $\rho_\mathrm{e}(z_i, k)$ is obtained by dividing the reduction in profit, $\Delta \hat{\pi}(z_i, k)$, by the vehicle SoC, $s_i(k)$, as shown in Equation~\eqref{eqn:value_of_charge_k}. If a vacant EV is not utilised in the initial local matching, its $\rho_\mathrm{e}(z_i, k)$ would be $0$.

    To compare the expected charging benefits between different matching possibilities, the future $\rho_\mathrm{e}$ is predicted and compared with the present value for each destination zone (i.e., around a passenger trip destination or a charger location). The value of $\rho_\mathrm{e}$ in the destination zone, $z_j$, at some time in the future, $k'$, requires a prediction of the passenger demand and the EV fleet SoC conditions. The passenger demand prediction can rely on statistical analysis of historical trip data and sample from a similar training dataset. On the other hand, while an accurate prediction of the fleet supply and its SoC level is challenging for longer time spans due to stochastic driver behaviours and uncertain matching outcomes, it can be assumed that a projection of the current vehicle actions (picking up and delivering passengers, recharging, etc.) yields reasonably good short-term market predictions that is sufficient for the vehicle to reach the destination zone.

    With the predicted local demand and supply conditions, a similar algorithm can be used to predict the future marginal value of charge, $\rho_\mathrm{e}(z_j, k')$. Instead of evaluating the profit loss without vehicle $i$, the destination zone considers the profit gain brought by the addition of vehicle $i$, as shown in Equations~\eqref{eqn:value_of_charge_k'_passenger} and \eqref{eqn:value_of_charge_k'_charger}. 
    \begin{align}
        \rho_\mathrm{e}(z_i, k) &= \frac{\Delta \hat{\pi}(z_i, k)}{s_i(k)} \label{eqn:value_of_charge_k}\\
        \hat{\rho}_\mathrm{e}^\mathrm{p}(z_j, k') &= \frac{\Delta \hat{\pi}(z_j, k')}{s_i(k) - Q_i^\mathrm{o}\left(t_{ij}^\mathrm{p} + t_j^\mathrm{o} \right)}, & k' = k + t_{ij}^\mathrm{p} + t_j^\mathrm{o}, ~\forall{j \in \mathcal{P}}\label{eqn:value_of_charge_k'_passenger}\\
        \hat{\rho}_\mathrm{e}^\mathrm{c}(z_j, k') &= \frac{\Delta \hat{\pi}(z_j, k')}{0.9 s_i^\mathrm{max}}, & k' = k + \Delta k_{ij}^\mathrm{c}, ~\forall{j \in \mathcal{C}}\label{eqn:value_of_charge_k'_charger}
    \end{align}

\nomenclature[D, 08]{$\rho_\mathrm{e}(z_i, k)$}{Marginal value of charge in the current zone around EV \nomunit{$\si{\$\per\kilo\watt\hour}$}}
\nomenclature[D, 09]{$\hat{\rho}^\mathrm{p}_\mathrm{e}(z_j, k')$}{Marginal value of charge in the future passenger zone \nomunit{$\si{\$\per\kilo\watt\hour}$}}
\nomenclature[D, 09]{$\hat{\rho}^\mathrm{c}_\mathrm{e}(z_j, k')$}{Marginal value of charge in the future charging station zone \nomunit{$\si{\$\per\kilo\watt\hour}$}}

    Equations~\eqref{eqn:charging_benefit_passenger} and \eqref{eqn:charging_benefit_charger} represent the expected charging benefit or loss for passenger and charging trips respectively, with their trip completion time for the future prediction. For example, if a vehicle is redundant in both zones, the expected charging benefit would be $0$ and does not affect the matching solution. If the vehicle has a higher marginal value of charge in the destination zone, the expected benefit is positive, and the overall matching benefit is increased to encourage such dispatches, vice versa.
    \begin{align}
        \Delta \rho_\mathrm{e}^\mathrm{p} &= \hat{\rho}_\mathrm{e}^\mathrm{p}(z_j, k + t_{ij}^\mathrm{p} + t_j^\mathrm{o}) - \rho_\mathrm{e}(z_i, k)\label{eqn:charging_benefit_passenger}\\
        \Delta \rho_\mathrm{e}^\mathrm{c} &= \hat{\rho}_\mathrm{e}^\mathrm{c}(z_j, k + \Delta k_{ij}^\mathrm{c}) - \rho_\mathrm{e}(z_i, k) \label{eqn:charging_benefit_charger}
    \end{align}

\subsection{Optimisation of expected matching benefit}
\label{sec:optimisation}
    % Objective and matching (with incentive) solution
    The optimal batch matching solution can be obtained by solving Equation~\ref{eqn:objective} for the set of vacant EVs ($\mathcal{V}$), waiting passengers ($\mathcal{P}$), and all chargers in the network ($\mathcal{C}$) at each matching interval. The expected benefit is not guaranteed for charging trips due to the probability of dispatch rejection (see driver compliance in Section~\ref{sec:behaviour_driver}). A discount between $0$ and $1$ as in Constraint~\eqref{const:incentive_bounds}, $r_{ij}$, is offered by the TNC to cover a fraction of driver charging fees, raising driver compliance at the expense of lower charging profit. The matching decision $x_{ij}$ is a binary variable as shown in Constraint~\eqref{const:binary}, indicating the optimal matchings. Some matching options are infeasible if the vehicle SoC is insufficient to complete a passenger trip as shown in Constraint~\eqref{const:passenger_feasibility} or reach the charger location as shown in Constraint~\eqref{const:charger_feasibility}. With an infinite search radius,\footnote{ Despite the infinite search radius, matchings would be cancelled by passengers if the pick-up time exceeds their patience levels. Unrealistically far pick-up trips will not happen.} the matching feasibility is only limited by SoC consumption. Last, Constraints~\eqref{const:col_sum} and \eqref{const:row_sum} ensure that at most one EV is dispatched to any passenger or charger at the same time.
    
    \begin{equation}
        \max_{r_{ij},x_{ij}} \sum_{i \in \mathcal{V}} \sum_{j \in \mathcal{\{P, C\}}}  C_{ij} x_{ij} \pi_{ij} \label{eqn:objective}
    \end{equation}
    \setcounter{equation}{10}
    \begin{subequations}
        \begin{align}
        \text{s.t.}\nonumber\\
        \pi_{ij}(k) &= (f - w) t_j^\mathrm{o} + \rho_\mathrm{t}^\mathrm{pm} t_j^\mathrm{m} - \rho_\mathrm{t}^\mathrm{pp} t_{ij}^\mathrm{p} + \Delta \rho_\mathrm{e}^\mathrm{p} Q_i^\mathrm{o}\left(t_{ij}^\mathrm{p} + t_j^\mathrm{o} \right) &\forall{j \in \mathcal{P}} \tag{\ref{eqn:benefit_passenger}} \\
        \pi_{ij}(k) &= e_{ij}(k) \Delta s_i - \rho_\mathrm{t}^\mathrm{c} \Delta k_{ij}^\mathrm{c} + \Delta \rho_\mathrm{e}^\mathrm{c} \Delta s_i  & \forall j \in \mathcal{C} \tag{\ref{eqn:benefit_charging}} \\
        C_{ij} &= 1 & \forall{j \in \mathcal{P}} \nonumber\\
        C_{ij} &= \ln\left(\gamma_i^1  + \gamma_i^2 r_{ij} + \gamma_i^3 \Delta k_{ij}^\mathrm{c} / 3600\right) &\forall{j \in \mathcal{C}} \tag{\ref{eqn:compliance}} \\
        r_{ij} &\in \left[0, 1\right] \label{const:incentive_bounds}\\
        x_{ij} &\in \{0, 1\} \label{const:binary}\\
        x_{ij} &= 0 \mbox{ if } s_i(k) < Q_i^\mathrm{o} (t^\mathrm{p}_{ij} + t_j^\mathrm{o}) &\forall{j \in \mathcal{P}} \label{const:passenger_feasibility}\\
        x_{ij} &= 0 \mbox{ if } s_i(k) < Q_i^\mathrm{o} t^\mathrm{c}_{ij} &\forall{j \in \mathcal{C}} \label{const:charger_feasibility}\\
        &\sum_{j \in \{\mathcal{P, C}\}} x_{ij} \leq 1 &\forall{i \in \mathcal{V}} \label{const:col_sum} \\
        &\sum_{i \in \mathcal{V}} x_{ij} \leq 1 &\forall{j \in \{\mathcal{P}, \mathcal{C}\}} \label{const:row_sum}
        \end{align}
    \end{subequations}

\subsection{Two-step optimisation}
    The proposed mixed-binary optimisation problem in Equation~\eqref{eqn:objective} can be split into two separate equivalent problems. \citet{jiao_incentivizing_2022} proved the equivalence in the context of shared and solo ride-hailing trips, where passengers receive discounted fares for shared trips. The first step is to determine the optimal incentive values for all potential charging trips, considering all matching pairs. The second step is to compute the optimal matching solution with the optimal incentive values from the previous step. 
    
    Equation~\eqref{eqn:objective_incentive} explicitly expresses the optimal incentive value in terms of the expected driver compliance and charging benefit. Note that the bound in Constraint~\eqref{const:incentive_bounds} still applies. Incentive optimisation is only required for charging trips.
    
    \begin{equation}
        r_{ij}^* = \underset{r_{ij}}{\arg \max} ~C_{ij} \pi_{ij} \qquad \forall {i \in \mathcal{V}}, {j \in \mathcal{C}} \label{eqn:objective_incentive}
    \end{equation}

    Given the optimal incentive values, the original optimisation problem in Equation~\eqref{eqn:objective} is transformed to Equation~\eqref{eqn:objective_matching}, which is a typical bipartite matching problem (or linear assignment problem) with weighted edges.

    \begin{equation}
        \max_{x_{ij}} \sum_{i \in \mathcal{V}} \sum_{j \in \mathcal{\{P, C\}}}  C_{ij}^* x_{ij} \pi_{ij}^* \label{eqn:objective_matching}
    \end{equation}

    To ensure equivalency, there is an assumption that the optimal incentive value for one charging trip is independent from another charging trip. It may seem unrealistic at first glance since if a charging station is assigned and occupied by one particular EV, the charging capacity is reduced for the other EVs. It affects the potential charging benefit and the resultant incentive values. However, matching constraint~\eqref{const:row_sum} should be considered, that at most one EV can be assigned to the charger in the batch matching process. In essence, the optimal incentive values are determined based on the charging capacities at the time of matching. Any changes in charging benefits due to matching would be considered in the next iteration of the batch matching problem, not affecting the current results.

\section{Passenger and Driver Behavioural Models}
\label{sec:behaviour}
    % Behaviour overview, what is the significance? Why necessary?
    In an on-demand mobility market, the short-term demand and supply conditions are highly dynamic due to complex user behaviours and trip choices. It is critical to recognise the effects of individual behaviours when evaluating the expected trip benefits for the TNC. We consider a range of relevant behavioural models such as the driver compliance as mentioned in prior sections, their shift scheduling under flexible working arrangements, and passenger's trip cancellation behaviours. Such behaviours are simulated in a testbed to reflect the performance of the proposed method. The proposed matching solution also considers passenger cancellation behaviours and driver exit probabilities when predicting the expected trip benefits. Passenger behaviours are governed by $2$ threshold parameters, while driver behaviours are governed by $6$ individual attribute parameters.

\subsection{Passenger trip cancellation choices}
\label{sec:behaviour_passenger}
\begin{figure}[htb!]
        \centering
        \includegraphics[width=0.83\textwidth]{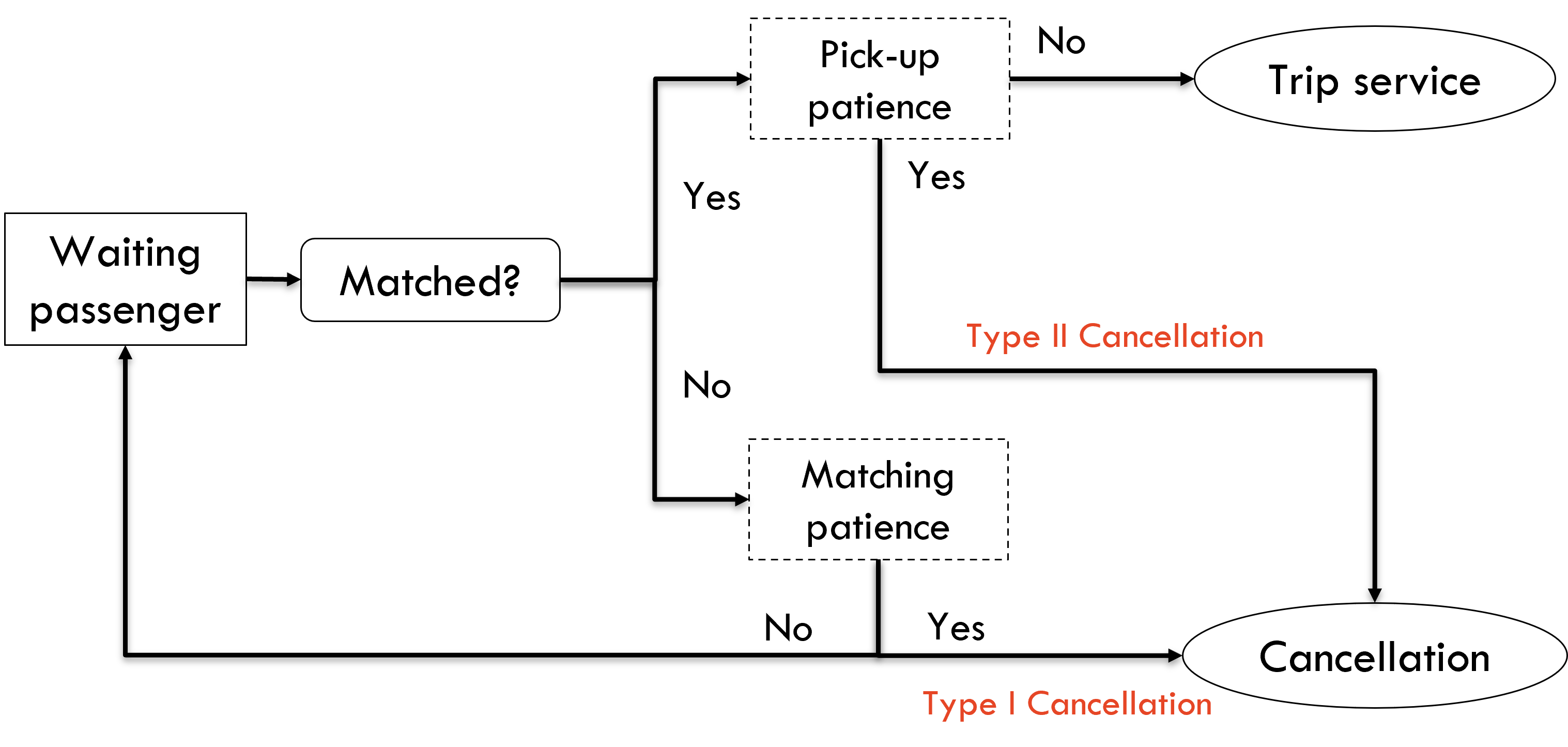}
        \caption{Decision-making of passengers trip acceptance and cancellation}
        \label{fig:passenger_decision}
    \end{figure}

    % Passenger request and Type I II cancellations
    As passenger $j$ hails for a trip from origin to destination at the request time via an online app, they may be matched at the next iteration of the batch matching algorithm. If the matching fails, the passenger compares the match waiting time $t_j^\mathrm{m}$ with a personal patience tolerance (e.g., around $1$ minute). Once the tolerance is exceeded, the passenger no longer waits for the next matching iteration and cancels their requests immediately (type I cancellation). Otherwise, they continue to wait for potential matchings. If the matching is successful, the TNC provides an accurate pick-up time (or estimated time of arrival), $t_{ij}^\mathrm{p}$, which the passenger compares against their personal waiting tolerance (e.g., around $7$ minutes) to determine if the matching outcome is acceptable. Passengers would cancel the trip if the matching results in unacceptable pick-up times (type II cancellation). Relevant passenger decisions are shown in Figure~\ref{fig:passenger_decision}.
    
\subsection{Driver compliance and work shift}
\label{sec:behaviour_driver}
\begin{figure}[htb!]
        \centering
        \includegraphics[width=0.83\textwidth]{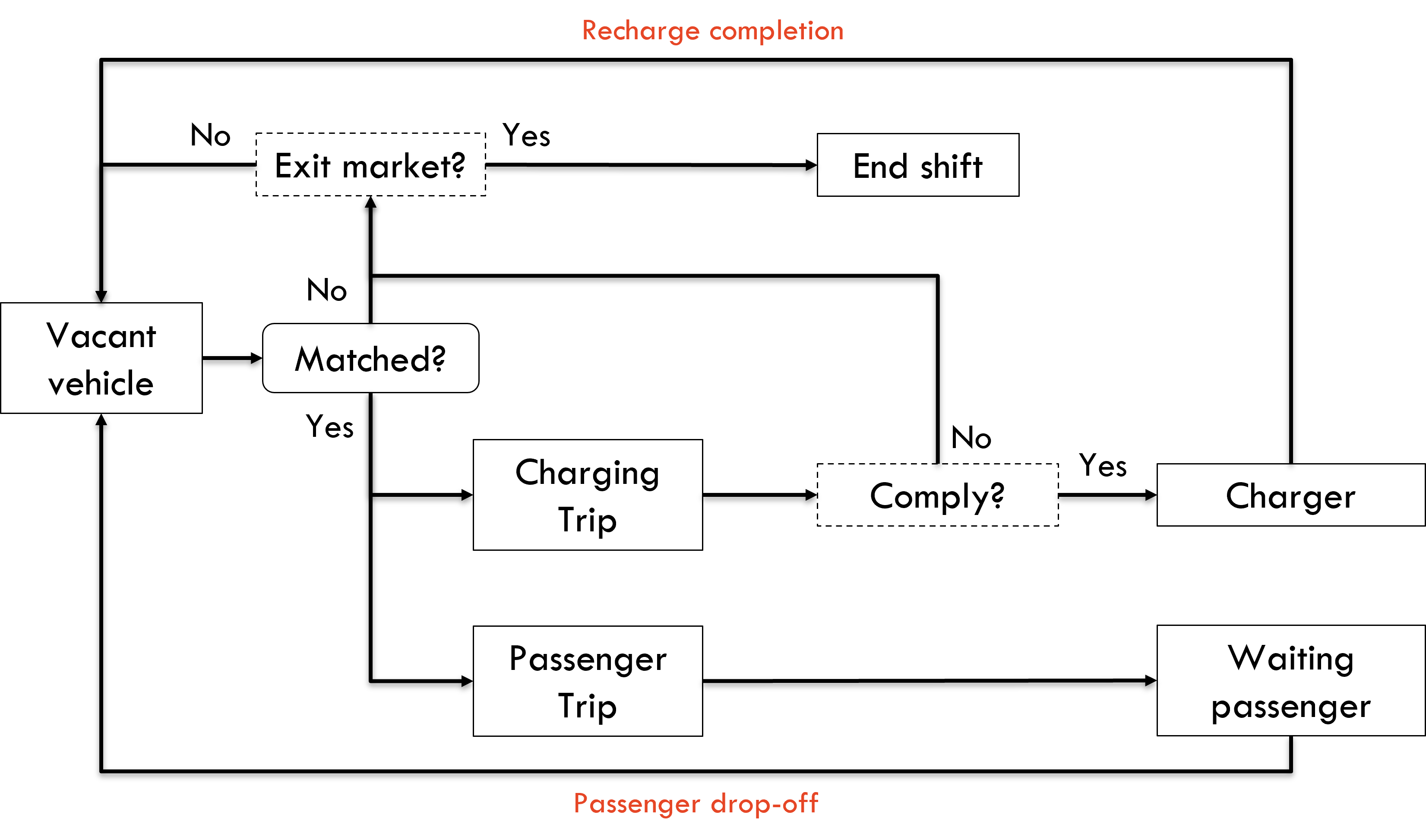}
        \caption{Decision-making of EV drivers}
        \label{fig:driver_decision}
    \end{figure}

    % Driver-passenger trip process
    Relevant driver decisions are shown in Figure~\ref{fig:driver_decision}. At each matching instance, vacant EV $i$ may be matched with a waiting passenger or EV charger $j$ in the network. Drivers are assumed to always accept passenger trips which directly contribute to their personal incomes. If passenger $j$ rejects the matching due to type II cancellation, vehicle $i$ returns to the vacant vehicle set $\mathcal{V}$ for the next iteration of matching. If the matching is not rejected, vehicle $i$ travels to the passenger trip origin for pick-up, and then to the destination for drop-off. 
    
    % Driver-charger trip process
    For a charging trip, the driver of vehicle $i$ decides whether to comply with the dispatching order based on the average incentive value, vehicle SoC, and the estimated charging time. If the driver decides to comply with the charging dispatch, vehicle $i$ travels to the matched charger $j$ and tops up the SoC to $90$\% of the maximum capacity\footnote{ In reality, the charging speed is non-linear and slows down at high SoC levels to protect the vehicle battery from degradation. Thus, this paper assumes EVs recharge to $90$\% capacity with a linear rate.} before re-joining set $\mathcal{V}$. If an EV is not matched with any passenger or charging station, or if the driver does not comply with the charging trip dispatch, the EV driver would consider whether to end their shift and leave the market. The exit decision recurs at most once every $1$ minutes for each driver.

\begin{figure}[hbt!]
        \centering
        \begin{subfigure}{0.54\textwidth}
            \includegraphics[width=\linewidth]{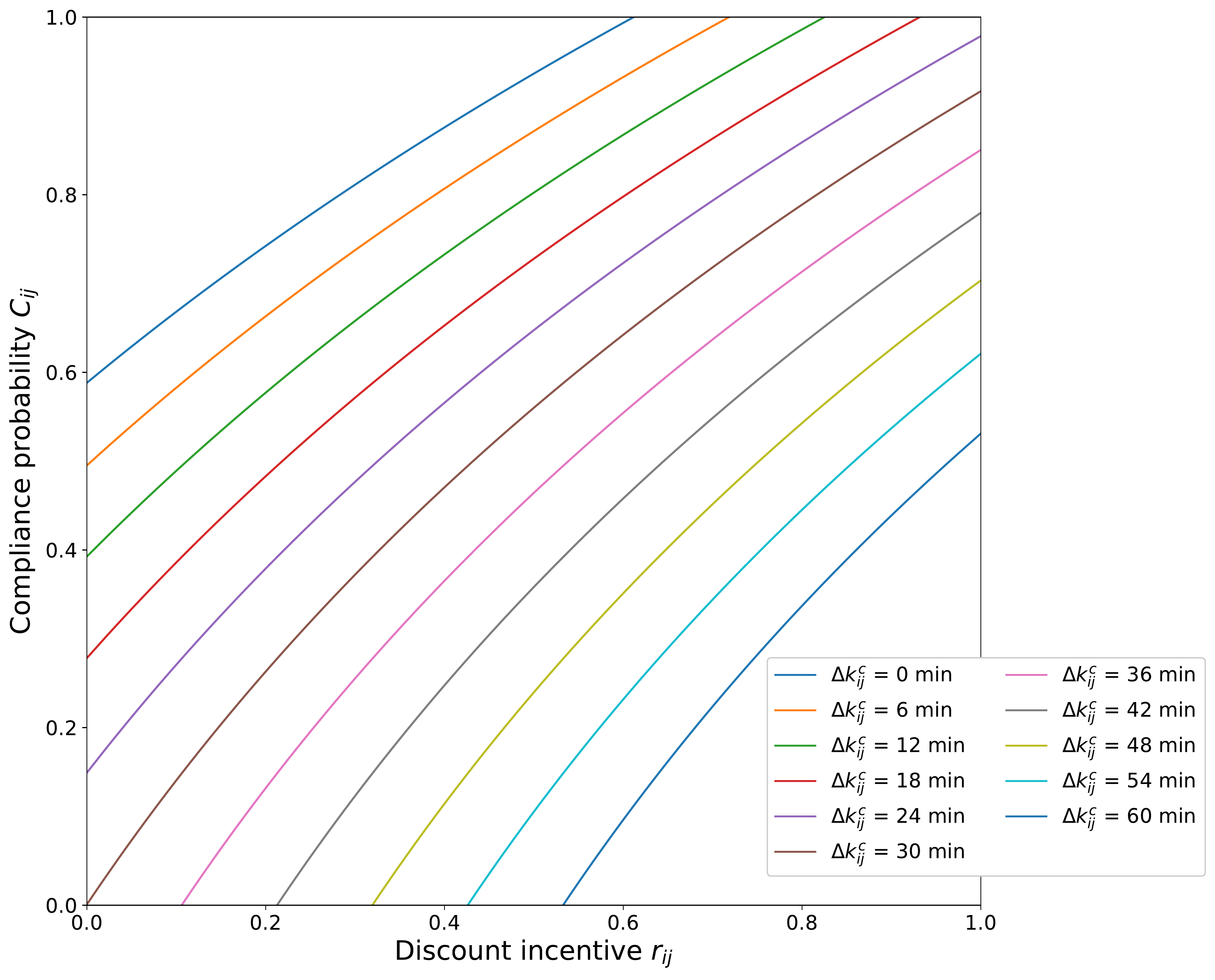}
            \caption{Compliance v.s. discount incentive}
        \end{subfigure}
        \hfill
        \begin{subfigure}{0.45\textwidth}
            \includegraphics[width=\linewidth]{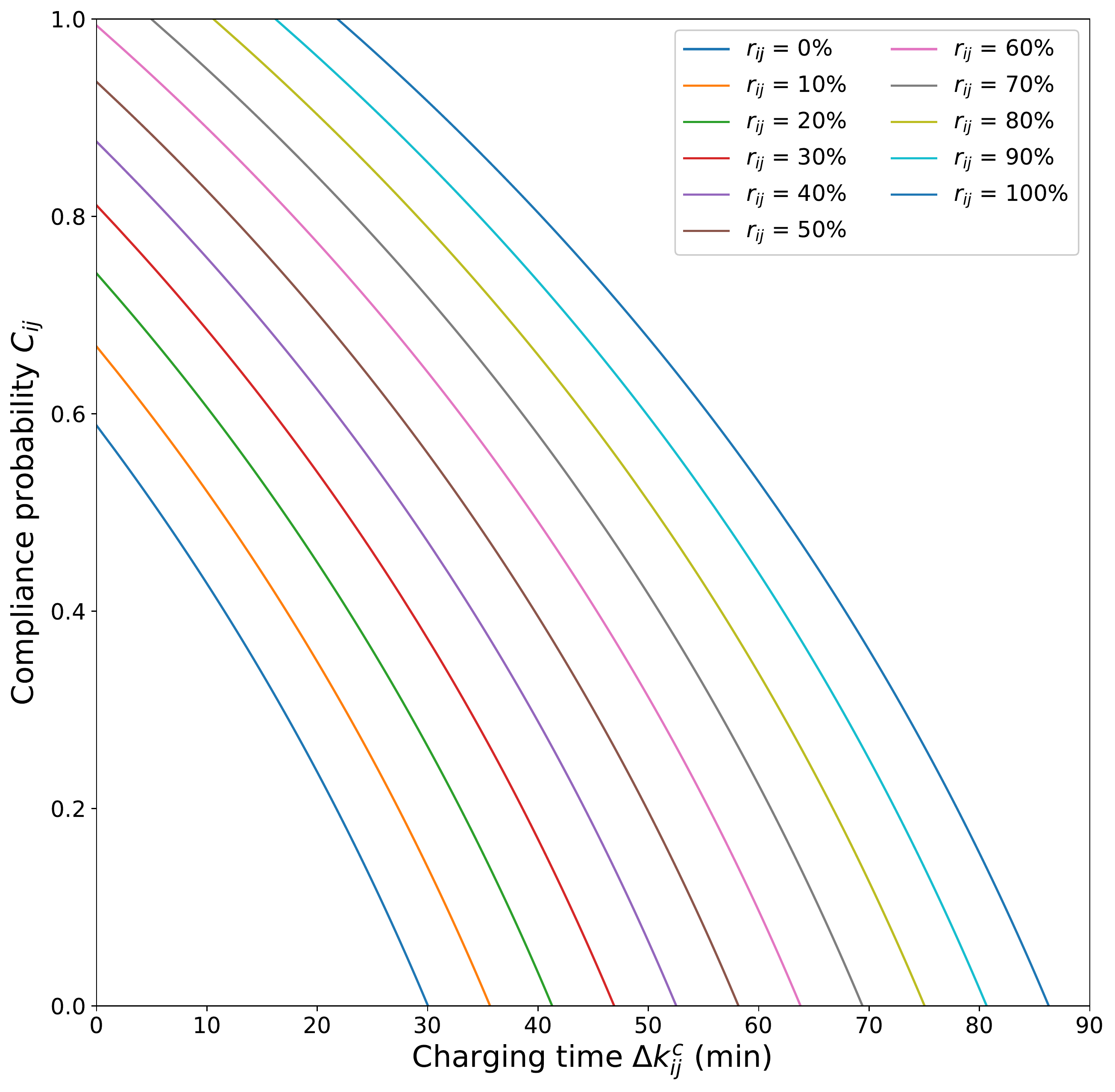}
            \caption{Compliance v.s. charging time}
        \end{subfigure}
    \caption{Example relationships between the driver compliance for the dispatched charging trips with (a) discount incentive offered by the TNC; and (b) charging times for a driver with attributes ($\gamma_i^1$, $\gamma_i^2$, and $\gamma_i^3$) of $1.8$, $1.5$, and $-1.6$ respectively. The compliance increases with the incentive and decreases with the charging time.}
    \label{fig:driver_compliance}
    \end{figure}

    % Compliance behaviour
    The behavioural studies on EV charging prices and incentivisation policies are scarce. \citet{wang_what_2020} investigates the new energy vehicle market in China and finds that lower operating cost is the second most important factor when a customer evaluates different purchase options. \citet{edwards_increasing_2002} discovers a logarithmic relationship between postal questionnaire response rates and the monetary incentive. Given the lack of empirical data, this paper estimates the dispatching compliance of driver $i$ with a bivariate logarithmic function as shown in Equation~\eqref{eqn:compliance}. The probability of compliance is dependent on the charging discount incentive (bounded between $0\%$ and $100\%$) and the estimated charging time (Equation~\ref{eqn:charging_time}). Individual attributes ($\gamma_i^1$, $\gamma_i^2$, and $\gamma_i^3$) are unique for each driver to represent their stochastic behaviours. The compliance is $1$ for passenger trips which are always accepted by the driver, and bounded between $[0, 1]$ for charging trips. The marginal effect of charging incentive diminishes with higher discounts. Figure~\ref{fig:driver_compliance} illustrate an example relationship between driver compliance and the discount value and vehicle charging time.
    \begin{align}\label{eqn:compliance}
          C_{ij} &= 1 & \forall{j \in \mathcal{P}} \nonumber\\
          C_{ij} &= \ln\left(\gamma_i^1  + \gamma_i^2 r_{ij} + \gamma_i^3 \Delta k_{ij}^\mathrm{c} / 3600\right) & \forall{j \in \mathcal{C}}
    \end{align}

\nomenclature[B, 05]{$C_{ij}$}{Compliance probability for driver $i$ to charge at charger $j$ \nomunit{-}}
\nomenclature[C, 06]{$\gamma_i^1, \gamma_i^2, \gamma_i^3$}{Driver compliance behavioural attributes \nomunit{-}}

    % Driver flexible working schedule
    Drivers in the on-demand mobility market have flexible working schedules. While the market participation decisions are usually formed by long-term habits, the exit decisions are affected by short-term market conditions and driver actions \citep{ashkrof_understanding_2020, ramezani_empirical_2022}. In this paper, drivers join the market at their pre-defined preferred starting times, which are sampled from a random distribution obtained by performing kernel density estimation on historical driver shift start data \cite{TLC2018}. Drivers may choose to end work after unsuccessful matchings, or when they decide not to comply with a charging trip dispatching order as shown in Figure~\ref{fig:driver_decision}.

\begin{figure}[hbt!]
        \centering
        \begin{subfigure}{0.47\textwidth}
            \includegraphics[width=\linewidth]{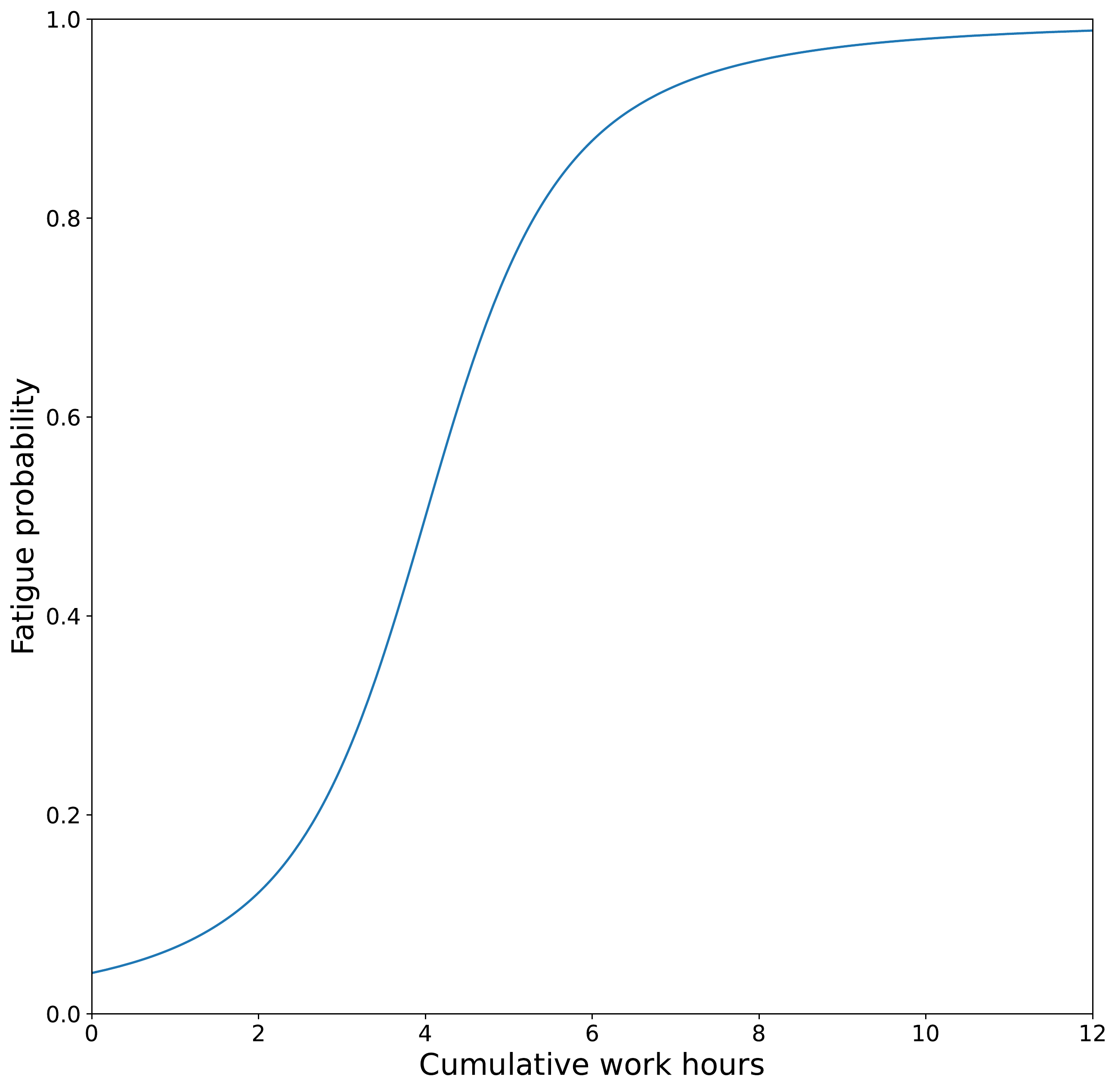}
            \caption{Fatigue probability}
        \end{subfigure}
        \hfill
        \begin{subfigure}{0.48\textwidth}
            \includegraphics[width=\linewidth]{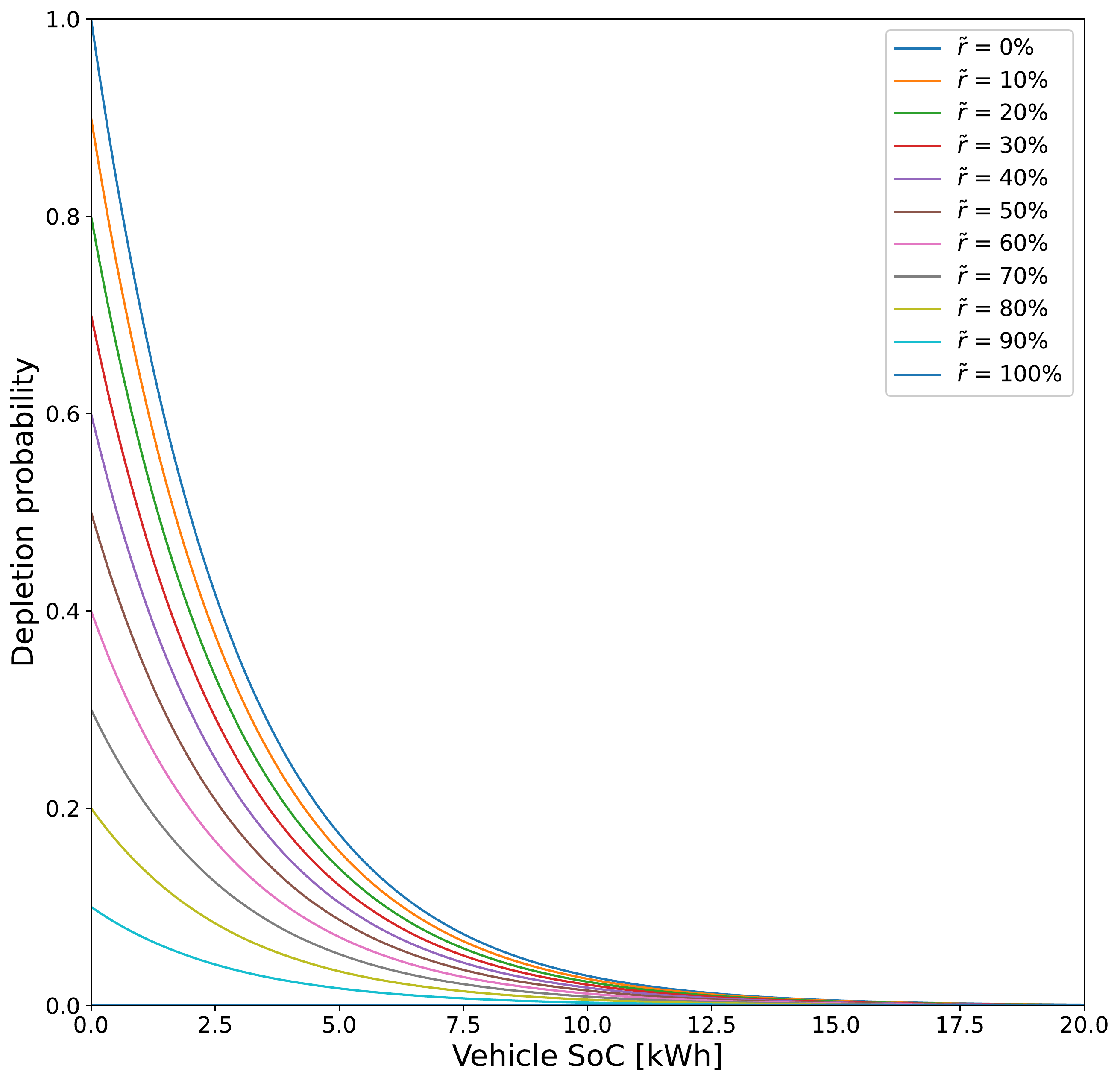}
            \caption{SoC depletion probability}
        \end{subfigure}
    \caption{Example exit probabilities due to (a) fatigue as a result of long work hours; and (b) vehicle SoC depletion at different market incentive levels for a certain EV driver with attributes ($\gamma_i^4$, $\gamma_i^5$, and $\gamma_i^6$) of $4$ [hours], $2$, and $0.2$ respectively.}
    \label{fig:exit_probability}
    \end{figure}

    % Exit probability function
    The exit probability is a joint probability of two possible cases: (1) when the driver's fatigue, represented by the cumulative work hours, $t_i^\mathrm{total}$, is too high so that the driver is physically uncomfortable to work any longer; and (2) when the vehicle SoC ($s_i$) is low, while the average TNC charging incentive ($\tilde{r}$) is also low. Considering the relatively high charging cost, a rational driver would leave the market to charge the EV at home. Equation~\eqref{eqn:exit_cases} specifies the probability of the two exit cases, and Equation~\ref{eqn:exit_probability} is the joint probability of market exit for driver $i$. Figure~\ref{fig:exit_probability} shows the two exit probabilities for a driver with selected attribute values.

    \begin{align}
        &\begin{cases}
            P_i(\text{fatigue}) &= 0.5 + \frac{t_i^\mathrm{total} - \gamma_i^4} {2 \sqrt{\gamma_i^5 + (t_i^\mathrm{total} - \gamma_i^4)^2}} \\
            P_i(\text{depletion}) &= \frac{1 - \tilde{r}}{\exp(\gamma_i^6 s_i)}\\
        \end{cases}\label{eqn:exit_cases}\\
        &P_i(\text{exit}) = P_i(\text{fatigue}) + P_i(\text{depletion}) - P_i(\text{fatigue}) P_i(\text{depletion}) \label{eqn:exit_probability}
    \end{align}

\nomenclature[C, 07]{$\gamma_i^4, \gamma_i^5, \gamma_i^6$}{Driver exit behavioural attributes \nomunit{$\si{\hour}$, -, -}}

\section{Simulation}
\label{sec:simulation}
    % Why simulation is necessary?
    To implement the proposed matching method and measure its performance in regard to both passenger trip service and fleet management, a detailed simulation is developed in Python as a testbed. The market is managed by a monopolistic TNC which is responsible for dispatching (or recommending) a fleet of EVs to the optimal trip matchings, which aims to maximise the TNC's expected profit. Passengers and drivers interact in a real-world road network and make individual decisions based on their own attributes and trip information provided by the TNC.

    % Road network
    The road network is derived from the topology of Manhattan, obtained from OpenStreetMap. After data cleaning and filtering, the network is represented as a strongly-connected directed graph within the boundaries of Manhattan. Coordinates are projected onto the nearest point in the network, at some distance along a directed edge. The resultant graph contains $4360$ nodes and $9537$ edges. This paper disregard microscopic traffic condition, e.g., vehicle acceleration and deceleration, lane-changing, and traffic signals. The network experiences no time-varying congestion, so that traffic flows of other vehicle types are not considered. EVs traverse each road at pre-defined road speeds, which are calibrated for each road based on historical trip records \citep{chen_decentralised_2021}.

\begin{figure}[hbt!]
        \centering
        \includegraphics[width=0.4\linewidth, trim={0 2cm 0 1cm}]{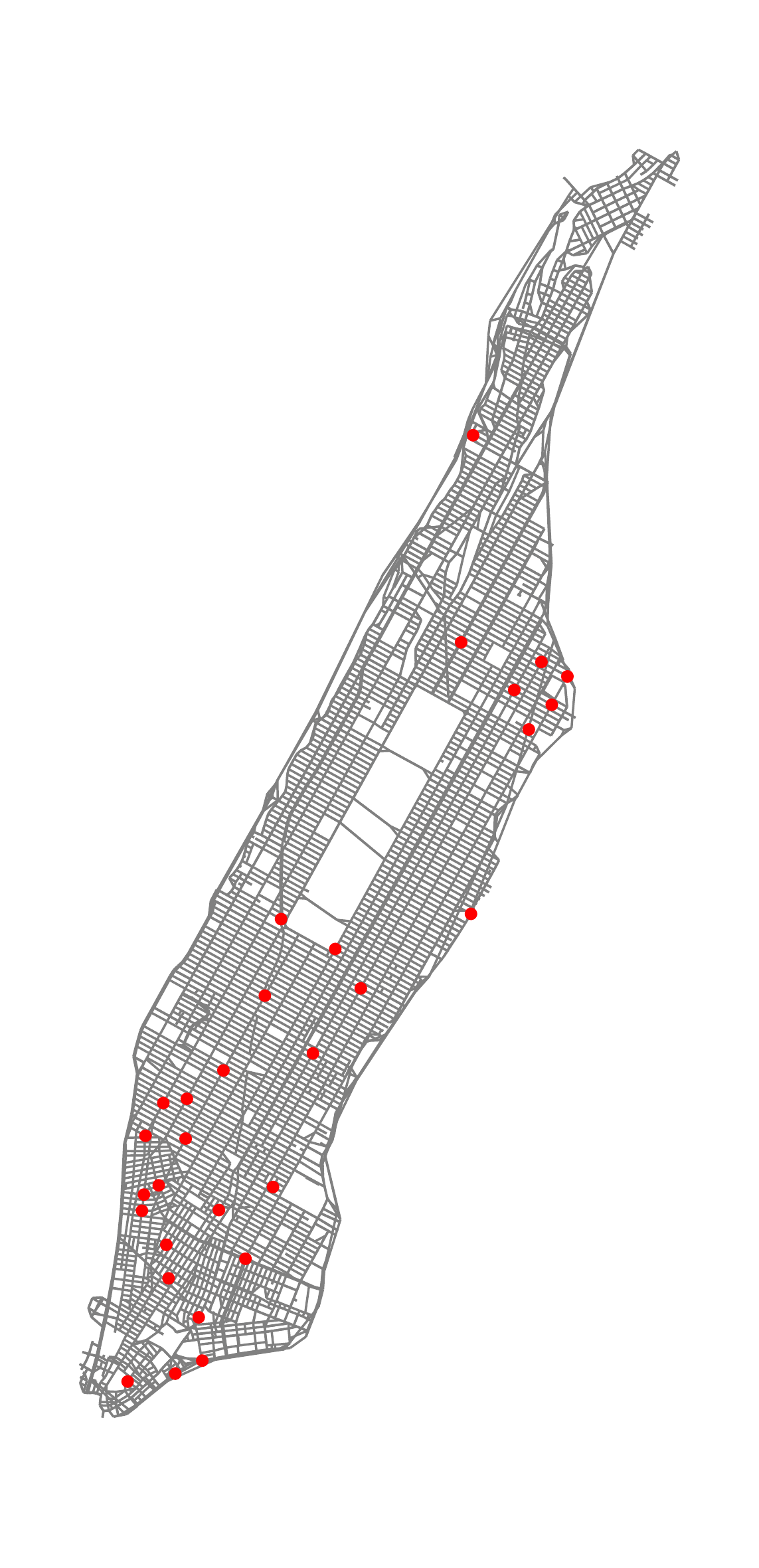}
        \caption{Charging station locations}
        \label{fig:charging_station}
    \end{figure}

    % Chargers
    A total of $30$ EV chargers are assumed installed at random intersections in the network. Multiple chargers at the same location are considered as a charging station. The setup assumes $6$ chargers (or charging piles) are installed at each charging station. A charger can only be occupied by one EV at a time. It is assumed that the TNC has access to accurate charging times, including the completion time of a charging EV and the arrival time of a reserved EV travelling to the charger. Although different charger types can be configured, the simulation setting assumes $120$ kW charging speed for all chargers. A charging vehicle is restored to $90$\% of its maximum SoC capacity at a constant rate. Locations of the charging stations are visualised in Figure~\ref{fig:charging_station}. The charging prices depend on the profit margin, infrastructure cost, and the time-of-use tariff prices ($e_1$, $e_2$, and $e_3(k)$) of the TNC. In the simulation, the profit margin is set as $\$0.1$/kWh and the infrastructure cost is $\$0.05$/kWh. The time-of-use tariff is $\$0.018$/kWh during off-peak hours from $00:00$ to $08:00$, and $\$0.255$/kWh during peak hours after $08:00$.

    % Demand simulation, test day = 1
    Trip demands for on-demand EV trip services are replicated from historical yellow taxi trip records \cite{TLC2016} in June 2016 with their respective trip origin and destination coordinates, using the historical taxi pick-up times as request times. Passengers pay \$$42$/hour to the TNC for their in-vehicle travel (service) time. The test days use 24-hour demand on 1 June 2016. Passengers would cancel the request if the waiting time till a successful match exceeds their patience threshold, or if the matched pick-up time is too long. The patience values are drawn from truncated normal distributions for each passenger, as summarised in Table~\ref{tab:patience}.

    \begin{table}[htb!]
        \footnotesize
        \centering
        \caption{Distribution statistics for passenger patience values}
        \begin{tabular}{l|ccc|c}
            \toprule
            & min & mean & max & standard deviation \\
            \midrule
            Matching patience & $30$ sec & $60$ sec & $90$ sec & $6$ sec \\
            \midrule
            Pick-up patience & $5$ min & $7.5$ min & $10$ min & $1$ min \\
            \bottomrule
        \end{tabular}
        \label{tab:patience}
    \end{table}

    % EV configuration
    For a straightforward comparison between different strategies, the total EV fleet size and driver shift start times are assumed to be known in advance, based on historical observations. A total of $28,000$ EV drivers join the market with random starting times on the test day, sampled from their historical distribution. The historical driver shift start times are obtained with kernel density estimation of the hourly yellow taxi driver shift data \citet{TLC2018}. An initial fleet size (e.g., $2000$ vehicles) is loaded at the beginning of simulation to represent remaining fleet from the previous day. In addition, the fleet is homogeneous, meaning that all EVs have the same maximum SoC capacity ($54$ kWh) and the same energy consumption rate per unit travel time ($6$ kW). EV drivers start their work with an initial SoC level uniformly distributed between $20$\% and $100$\%, at a random location in the network. 
    
    A vacant EV can be matched with a waiting passenger or a charger. The matched or assigned vehicle then travels to the location of the passenger or charger to complete a trip service or vehicle recharging before becoming vacant again, if the driver decides not to exit the market. Drivers do not reposition or cruise in the network. Vacant EVs remain stationary upon passenger drop-offs or after completing their charging process. Each driver receives \$$30$/hour when the vehicle is occupied for passenger trip services. EV drivers may choose to exit the market and end their work shifts based on the matching results and compliance decisions as explained in Section~\ref{sec:behaviour_driver}. The individual compliance and exit behaviour attributes are drawn from truncated normal distributions as described in Table~\ref{tab:driver_behaviour_parameters}.

    \begin{table}[htb!]
        \footnotesize
        \centering
        \caption{Distribution statistics for driver compliance behavioural attributes}
        \begin{tabular}{l|ccc|c}
            \toprule
            & min & mean & max & standard deviation \\
            \midrule
            $\gamma_i^1$ & 1.5 & 1.8 & 2.1 & 0.1 \\
            \midrule
            $\gamma_i^2$ & 1.2 & 1.5 & 1.8 & 0.1 \\
            \midrule
            $\gamma_i^3$ & -1.9 & -1.6 & -1.3 & 0.1 \\
            \midrule
            $\gamma_i^4$ & 0.5 hour & 4 hour & 8 hours & 1 hour \\
            \midrule
            $\gamma_i^5$ & 0.8 & 2 & 4.2 & 0.4 \\
            \midrule
            $\gamma_i^6$ & 0.05 & 0.2 & 0.35 & 0.05 \\
            \bottomrule
        \end{tabular}
        \label{tab:driver_behaviour_parameters}
    \end{table}

\subsection{Benchmark strategies}
\label{sec:benchmark}
    To measure the performance of the proposed method, several benchmark strategies are designed to show the significance of the incentivisation policy, and the advantage of optimising the expected benefit which considers predictive market conditions instead of myopic profit maximisation. By considering the marginal value of charge, the proposed matching method is able to proactively advise vacant EVs to charge before demand surges.

    The first set of benchmark strategies disregard the TNC's profit when matching vacant EVs with waiting passengers. The objective is to minimise the total pick-up time for the passengers as shown in Equation~\ref{eqn:objective_min_tt}.
    \begin{equation}\label{eqn:objective_min_tt}
        \max_{x_{ij}} \sum_{i \in \mathcal{V}} \sum_{j \in \mathcal{P}}  \frac{x_{ij}}{t_{ij}^\mathrm{p}}
    \end{equation}

    The TNC does not attempt to manage the fleet SoC balance with respect to the predicted demand variations. Instead, a reactive charging policy is used such that drivers are reminded to charge their EVs at the closest (including the queuing time) charging station once the SoC drops below $10\%$ of the maximum capacity. Two scenarios of the benchmark are tested: one without any incentivisation such that the TNC obtains charging profits from EV drivers who would show relatively low dispatch compliance, and the other with full incentivisation (free charging for drivers) such that the TNC operates the charging stations at a loss to encourage higher driver compliance.

\section{Preliminary Results}
\label{sec:result}
    The market performance results are obtained from a $24$-hour simulation based on historical demand on 1 June 2016. Table~\ref{tab:benckmark_results} summarises key market performance indicators for benchmark scenarios without any incentivisation and free charging. The results show that the TNC suffers from a lower total profit when EV charging is offered for free.

    \begin{table}[hbt!]
    \caption{Market performance results for the benchmark strategy}
        \footnotesize
        \centering
        \begin{tabular}{l|l|r|r}
            \toprule
            && \textbf{No incentive} & \textbf{Free charging} \\
            \midrule
            \multirow{3}{*}{\textbf{Supply}} &
              Mean driver income* & \$52.33 & \$56.61 \\
            & Mean shift length (hour) & 2.55 & 2.70 \\
            & Mean EV initial SoC & 32.4 kWh & 32.4 kWh \\
            & Mean EV final mean SoC & 26.0 kWh & 29.4 kWh \\
            & \multicolumn{3}{l}{* Personal income after deducting charging fees.}\\
            \midrule
            \multirow{3}{*}{\textbf{Demand}} &
            Served passengers & 212512 (69.5\%) & 217157 (71.0\%)\\
            & \underline{\textbf{Cancellations}} & 93464 & 88819 \\
            & Type I Cancellation & 49223 & 42875 \\
            & Type II Cancellation & 44241 & 45944 \\
            & Mean matching time (s) & 11.9 & 11.1 \\
            & Mean pick-up time (s) & 221.8 & 223.0 \\
            \midrule
            \multirow{3}{*}{\textbf{TNC}}
            & \underline{\textbf{Number of chargings}} & 4593 & 6910 \\
            & Off-peak & 172 & 385 \\
            & Peak & 4421 & 6525 \\
            & \underline{\textbf{Charging profit}} & \$20745.59 & -\$91696.32 \\
            & Off-peak & \$771.61 & -\$1177.96 \\
            & Peak & \$19973.98 & -\$90518.36 \\
            & \underline{\textbf{Profit}} && \\
            & Trip profit & \$619196.16 & \$634001.63 \\
            & \textbf{Monetary profit} & \textbf{\$639941.75} & \textbf{\$542305.31} \\
            \bottomrule
        \end{tabular}        
        \label{tab:benckmark_results}
    \end{table}

\begin{figure}[hbt!]
        \centering
        \includegraphics[width=0.8\linewidth]{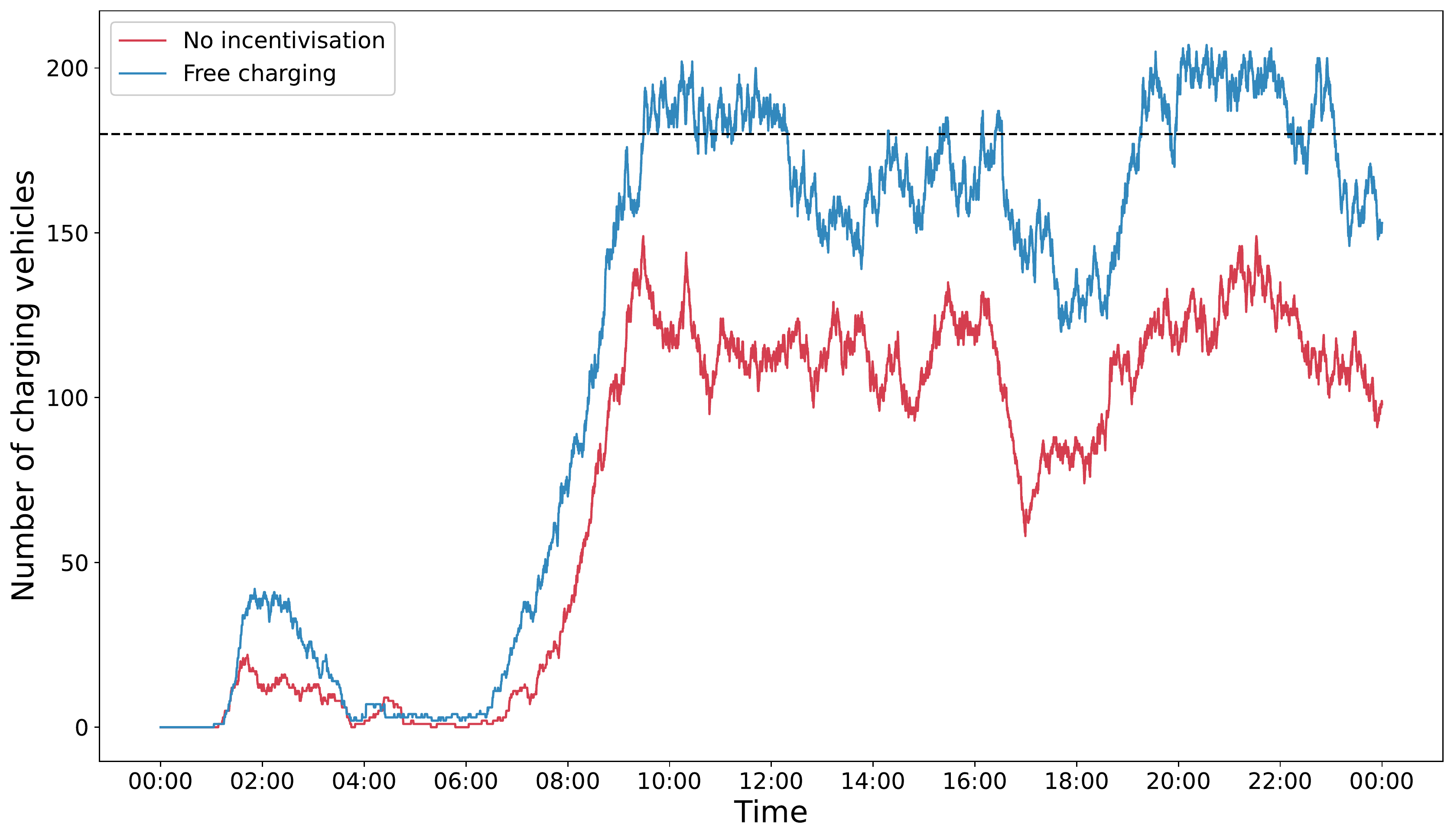}
        \caption{The number of charging vehicles. Note that queuing vehicles are also counted as charging vehicles, but only a total of $180$ chargers (charging piles), indicated by the dash line, are available in the network.}
        \label{fig:timeseries_nc}
    \end{figure}

    As expected, the main difference between the two scenarios is the number of EV charging and the associated costs. As evident in Figure~\ref{fig:timeseries_nc}, the number of complied charging trips is constantly higher when the charging is provided for free. It is worth pointing out that the off-peak and peak hour prices contribute to the disproportionate net profit disparities between the two scenarios. While the profit to charging vehicle ratio is about $\$9.15$ per vehicle during off-peak hours, this ratio rises $\$52.52$ per vehicle during peak hours, significantly reducing the economical efficiency of charging incentivisations. The benchmark results show the need of a well planned incentivisation scheme which significantly impacts the TNC's profit.

\begin{figure}[hbt!]
        \centering
        \begin{subfigure}{0.75\textwidth}
            \includegraphics[width=\linewidth]{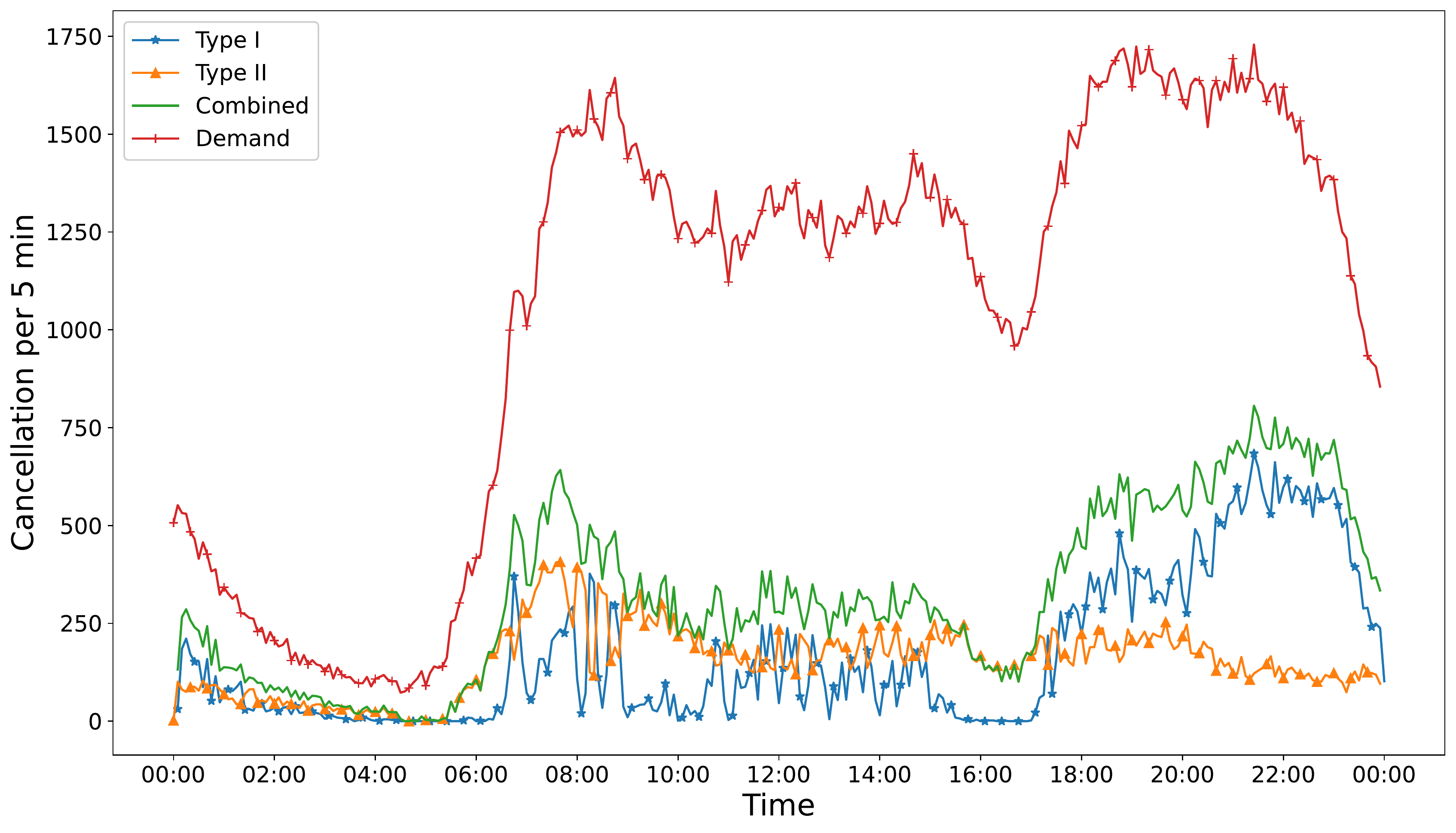}
            \caption{No incentivisation}
        \end{subfigure}
        % \hfill
        \begin{subfigure}{0.75\textwidth}
            \includegraphics[width=\linewidth]{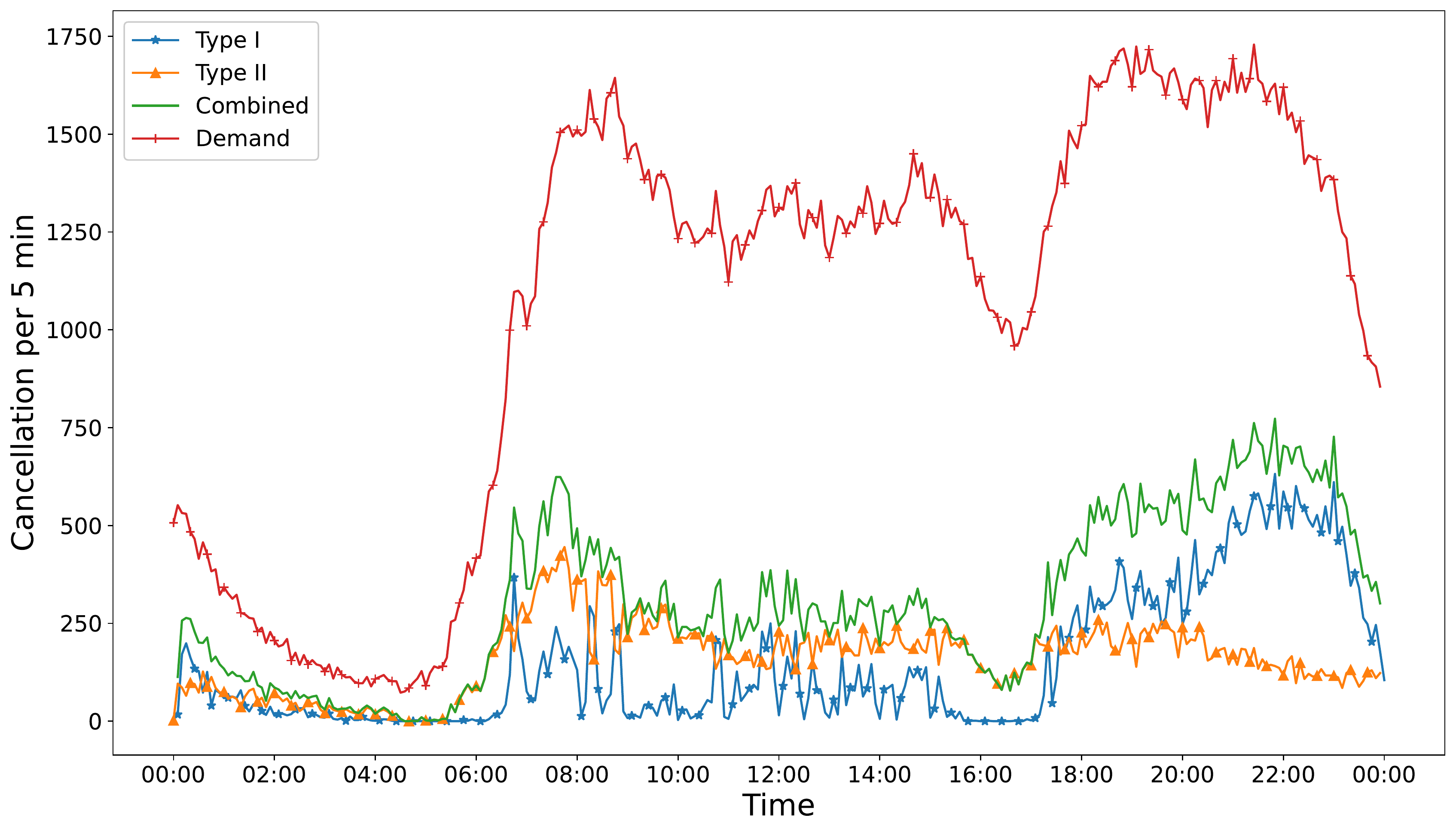}
            \caption{Free charging}
        \end{subfigure}
    \caption{The number of trip cancellations aggregated per $5$ min for the benchmark strategy (a) without incentivisation; and (b) with free charging.}
    \label{fig:timeseries_cancellation}
    \end{figure}

    Figure~\ref{fig:timeseries_cancellation} shows the number of cancelled trip requests. The two scenarios result in similar cancellation patterns even though the overall number of type I cancellations (inability to to matched with vacant EVs within the patience tolerance) is lower for the case with free charging. In both scenarios, type II cancellation outnumbers type I cancellation in the morning from $07:00$ to $11:00$. The opposite is obversed in the afternoon where type I cancellation becomes dominant after $18:00$.
    
\section{Summary and Future Work}
\label{sec:conclusion}
    This paper presents a centralised matching method which incorporates the charge scheduling problem within its optimisation framework. A metric called the marginal value of charge is proposed to quantify the predictive benefit of additional vehicle SoC in a local market around an EV. This approach combines the traditional dispatching problem with fleet SoC management in a dynamic environment with complex supply-demand interactions. Furthermore, this paper also designs an incentivisation policy to optimise human drivers' compliance behaviours for dispatching orders to charge their EVs. In reality, the stochastic driver behaviours would undermine the optimality of matching solutions. By offering charging discount incentives, the proposed method is able to increase the success rates of dispatching, achieving more profitable outcomes.

    To implement the proposed method and complete this paper, the quality of supply prediction as described in Section~\ref{sec:value_of_charge} needs to be tested. More benchmark results from multiple simulation runs will be included to demonstrate the effectiveness of the proposed method.

\newpage
\appendix
\nolinenumbers
\printnomenclature
%\linenumbers
    
\newpage
\bibliographystyle{elsarticle-harv_custom}
%\bibliography{probabilistic_EV.bib}
\bibliography{main.bib}





% \end{thebibliography}
    
\end{document}